\documentclass{webofc}
\usepackage[varg]{txfonts}   

\usepackage{natbib}

\usepackage{hyperref}
\usepackage{ragged2e}
\usepackage{lineno}
\usepackage{comment}
\begin{document}

\title{Building a Distributed Computing System for LDMX}
\subtitle{Challenges of creating and operating a lightweight e-infrastructure for small-to-medium size accelerator experiments}

\author{\firstname{Lene Kristian} \lastname{Bryngemark}\inst{1}\fnsep\thanks{\email{lkbryng@stanford.edu}; Corresponding author} \and
        \firstname{David} \lastname{Cameron}\inst{2}
        \and
        \firstname{Valentina} \lastname{Dutta}\inst{3} 
        \and 
        \firstname{Thomas} \lastname{Eichlersmith}\inst{4}
        \and
        \firstname{Balazs} \lastname{Konya}\inst{5}
        \and
        \firstname{Omar} \lastname{Moreno}\inst{6}
        \and  
        \firstname{Geoffrey} \lastname{Mullier}\inst{5}
        \and 
        \firstname{Florido} \lastname{Paganelli}\inst{5}
        \and
        \firstname{Ruth}   \lastname{Pöttgen}\inst{5}
        \and
        \firstname{Fuzzy} \lastname{Rogers}\inst{3}
        \and  
        \firstname{Andrii} \lastname{Salnikov}\inst{5}
        \and 
        \firstname{Paul} \lastname{Weakliem}\inst{3}
}

\institute{ 
    Stanford University, 450 Jane Stanford Way, Stanford, CA 94305, USA  
\and
    University of Oslo, P.b.\ 1048 Blindern, 0316 Oslo, Norway
\and 
    UC Santa Barbara, Santa Barbara, CA 93106, USA
\and
    University of Minnesota, Minneapolis, Minnesota, 55455, USA 
\and
    Lund University, BOX 118, S - 221 00 Lund, Sweden    
\and 
    SLAC, 2575 Sand Hill Rd, Menlo Park, CA 94025, USA
    }

\abstract{

Particle physics experiments rely extensively on computing and data services, making e-infrastructure an integral part of the research collaboration. Constructing and operating distributed computing can however be challenging for a smaller-scale collaboration.

The Light Dark Matter eXperiment (LDMX) is a planned small-scale accelerator-based experiment to search for dark matter in the sub-GeV mass region. Finalizing the design of the detector relies on Monte-Carlo simulation of expected physics processes. 
A distributed computing pilot project was proposed to better utilize available resources at the collaborating institutes, and to improve scalability and reproducibility.

This paper outlines the chosen lightweight distributed solution, presenting requirements, the component integration steps, and the experiences using a pilot system for tests with large-scale simulations. The system leverages existing technologies wherever possible, minimizing the need for software development, and deploys only non-intrusive components at the participating sites. The pilot proved that integrating existing components can dramatically reduce the effort needed to build and operate a distributed e-infrastructure,  making it attainable even for smaller research collaborations.
}
\maketitle
\section{Introduction}
\label{intro}
Particle and nuclear physics experiments rely extensively on e-infrastructures (various computing and data services) for generating, processing, analyzing, and storing large datasets.
While federating limited resources among collaborators enables a robust e-infrastructure, the nature and size of available resources for construction and operation depend largely on the scale of the collaboration. Large experimental collaborations can afford to develop, build and operate rather complex customized infrastructure utilizing large-scale dedicated resources, such as the Worldwide LHC Computing Grid (WLCG)~\cite{wlcg, wlcg2}. In contrast, building and operating a distributed e-infrastructure can be a substantial challenge for small to medium size collaborations, which often lack developers, operating teams, and access to significant dedicated data centers. 

The Light Dark Matter eXperiment (LDMX)~\cite{ldmx} is a planned small-scale accelerator-based experiment to search for dark matter. 
The nature of dark matter is one of the most pressing outstanding questions in particle physics. 
LDMX is a fixed-target missing momentum experiment currently being designed to probe the sub-GeV mass range, an attractive experimental regime for so-called thermal models with light mediators. 
It will hermetically measure the interactions of an incoming electron beam, provided by the upgraded Linac Coherent Light Source~\cite{LCLS2} linear accelerator at SLAC, with a target. Reaching the required unprecedented sensitivity translates both to high integrated luminosity (up to $10^{16}$ electrons on target) and to excellent knowledge of expected backgrounds from known physics processes.

With its roughly 30 collaborators across nine institutes, LDMX is a typical example of a small to medium size research collaboration. 
When it comes to e-infrastructure needs, it is no different from many other particle physics experiments. For the design of the LDMX apparatus, detailed simulations of a large number of events (incoming electrons) are required. To obtain a sufficient sample of the rarest background processes, the total number of simulated events needed is similar to the final number of expected events recorded in detector data. In the first phase of data taking, LDMX is expected to need on the order of 10PB of storage for collected data and simulations, and CPU equivalent of 1200 cores. 

Experiments have options to consider before deciding on final e-infrastructure. The possibilities range from the completely centralized solution (all the data are stored and all the computing tasks are performed at a large central data center at one of the partner sites) to various levels of distributed systems. Both in the case of the centralized system and the distributed approach there should be a well-defined, low-barrier mechanism for collaborating partners to integrate into the infrastructure and contribute to, and use, the resource pool. During 2020 the LDMX Collaboration board decided to launch a pilot project that would build a lightweight distributed computing system (LDCS) prototype for running simulation tasks. 
With such a system, LDMX could leverage and efficiently share available computing and storage resources at collaborating institutes. 
Distributed IT infrastructures are often considered complex solutions consisting of numerous additional services on top of the local system; an approach mostly suitable for large scientific collaborations. 
Two of the main goals of the pilot were to prove that a lightweight system can be created using existing technologies without the need for extensive new development, and to understand the effort required to operate (run) a distributed computational infrastructure as compared to a centralized system.

In what follows, we present the philosophy behind the design of the proposed distributed system, introduce the architecture and system components chosen, then give details of the experience gained during the large scale simulation test campaigns and outline future work.

\section{The LDMX Distributed Computing System}
\label{LDCS}

Initially, LDMX simulations were run with manual submission to the batch system at SLAC, the proposed location for the detector. SLAC is the primary data storage hub for the LDMX experiment. Any data not produced at SLAC would have to be transferred there.

LDCS was proposed as a distributed computing infrastructure alternative to a fully centralized data center. It is set up between four computing centers accessible to the LDMX collaboration: three located in California, USA, and one in Lund, Sweden. Each center runs a computing cluster (with a SLURM or HTCondor batch system) and has storage space available on a shared file system. Some centers provide external access to the data (e.g.\ via GridFTP protocol). The main objective of the distributed system pilot was to test if a small experiment such as LDMX could build and reliably operate a lightweight streamlined distributed system using existing technologies, with only a minimal effort invested in component integration.
The proposed design was based on the following considerations:

\begin{itemize}
\item The software layer implementing the distributed system, i.e.\ the middleware connecting the sites, should be lightweight, easy to deploy and operate, and preferably have a proven track record in particle physics.

\item A handful of sites would constitute the core LDMX computing infrastructure. In addition to those “core sites”, the system should allow additional, resource-limited sites to easily join in the longer term. The estimated number of such sites is on the order of ten. 

\item Core sites, LDMX Computing and Storage Centers (LCSCs), run the large scale simulations and store the generated results. An LCSC is assumed to have sufficiently large computing capacity (a large cluster with many cores) and a matching data storage capability. An LCSC is expected to offer longer term storage capacity in addition to being a large simulation processing unit with available matching temporary storage space.

\item Execution of computational tasks would be managed via a central production gateway, the job manager. This central service would interact with the lightweight middleware component deployed on every LCSC cluster.

\item The user administration and authentication layer would be kept centralized, i.e.\ creation of individual accounts on each LCSC site should not be necessary. LDMX users would access the data through authentication via a singular LCSC site, which hosts the “Collaboration user management and authentication service”. User-level authentication would be required against the central production gateway, while the gateway itself would authenticate  against the LCSC clusters via e.g.\ X.509 proxy credentials. All collaboration tasks running on the LCSC resources would be executed under “LDMX proxies”.

\item Job brokering would be data-driven, i.e.\ computational tasks should be dispatched to the LCSC that already stores any input data needed. In case no available site has a replica of the input data, the middleware component would take care of downloading the data. Later a dedicated data transfer service could be integrated to manage cross-site data transfers.

\item Generated data (e.g.\ simulation output) should always be reliably transferred to a primary storage with the option of keeping a local copy. The data transfer functionality would be offered by the middleware and require no manual intervention from the researchers.  In addition, a metadata record describing the generated data would be created and registered in a data catalog which also tracks data location. 

\end{itemize}

\subsection{The LDCS Architecture}
\label{arch}

The guiding philosophy of LDCS is to limit necessary site development and to integrate as few, already existing, components as possible. Leveraging existing technologies led to combining several components already deployed within the ATLAS experiment~\cite{atlas} at CERN:

\begin{itemize}
\item An LCSC site consists of a “traditional” batch cluster connected to a site-local storage system available to the compute nodes. The Advanced Resource Connector (ARC, see Sec.~\ref{sec:arc}) is deployed on the site and acts as the middleware layer, or  gateway, to the cluster and local storage. ARC provides the local layer for the data-driven task management. 
\item The LDCS central management layer consists of the ARC Control Tower (aCT, see Sec.~\ref{sec:act}) workflow management service and the LDMX data catalog (Rucio, Sec.~\ref{sec:rucio}).
\item A user-facing layer (yet-to-be developed) will provide indirect access to LCSC sites via the aCT component.
\end{itemize}

\begin{figure}[ht]
\centering
\includegraphics[width=11cm,clip]{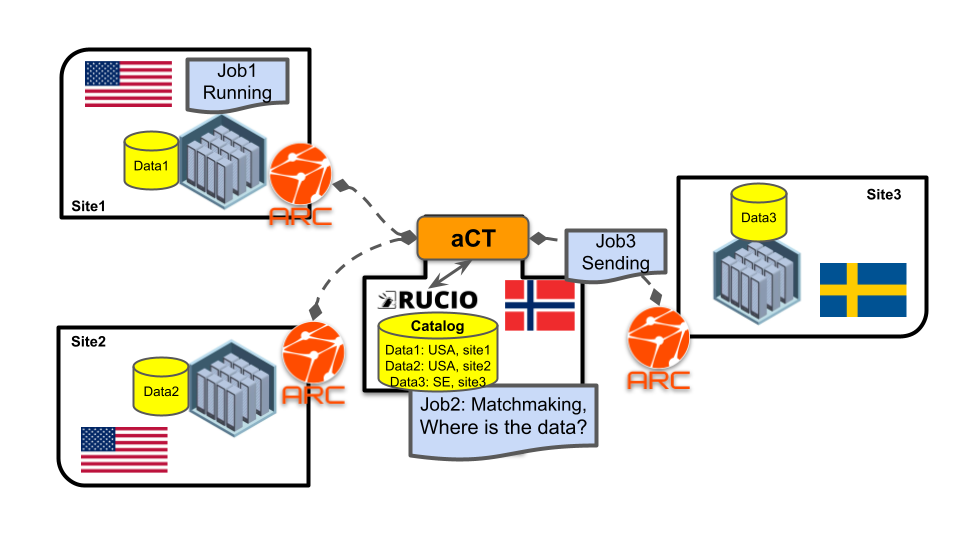}
\caption{The LDMX distributed computing system components. The LCSC sites are federated via the ARC middleware and the two central services aCT and Rucio.}
\label{fig:ldmx-arch}
\end{figure}

Figure~\ref{fig:ldmx-arch} illustrates the LDCS architecture. The components are described below.

\subsection{Advanced Resource Connector}
\label{sec:arc}

The Advanced Resource Connector (ARC)~\cite{arc} is developed by the NorduGrid collaboration and has been used for almost 20 years as a means of providing computing resources to WLCG experiments. Its lightweight design makes it an attractive choice for data centers who want to add a secure gateway to allow batch job submission from the outside. It also provides sophisticated data handling tools which relieve the jobs themselves from dealing with data transfer. Finally, the fact that ARC supports the batch systems used at all four centers made it the obvious choice for connecting them to the LDCS system.

ARC's main service is the ARC Resource-coupled EXecution service (A-REX) which runs on top of a computing center's batch system and allows users authenticated via X.509 credentials to submit and run jobs on the batch system. Each center in LDCS runs an A-REX service and accepts jobs submitted by the LDCS production certificate. When the job finishes on the batch system, the output is uploaded by A-REX to the selected LCSC storage. Having A-REX perform the data handling instead of the job itself brings several advantages: batch system CPUs are not idle while the job uploads data, controlled data staging by one service avoids chaotic transfer which may overload the remote storage, and if the remote storage is temporarily offline A-REX can automatically retry until the transfer succeeds.

\subsection{ARC Control Tower}
\label{sec:act}

ARC Control Tower (aCT)~\cite{act} is a service used by the ATLAS experiment to submit and manage jobs on ARC sites in the WLCG. It consists of two parts: an application-facing side and a resource-facing side. The application-facing side either pulls work from an experiment-specific workflow management system or generates work from tasks defined in aCT itself. An LDMX-specific component was developed to handle the management of LDMX tasks. The LDMX user who is the production manager defines a batch of jobs using a plain-text configuration file, detailing, for example, walltime needed, number of jobs, what software and simulation steering files to run, and submits the job request to aCT. aCT  creates the required jobs, specifying the resources required for memory and walltime as well as any input files, and passes the job to the resource-facing side. This side manages the communication with the ARC Computing Elements at the sites, submitting new jobs and querying their status.

aCT employs an intelligent brokering algorithm to decide when and where to submit jobs. If input data are required then it can query Rucio (see Sec.~\ref{sec:rucio}) for the location of the data, and send the job to where the data are, in order to avoid data transfers between sites. If no input data are required then the job will be sent to the site with the fewest queued jobs. A threshold on the absolute and relative queue depth at each site can be tuned in the aCT configuration.

\subsection{Rucio}
\label{sec:rucio}
Rucio~\cite{rucio} is a distributed data management system initially developed by ATLAS but now widely adopted by many scientific communities in particle physics and beyond. It consists of a data catalog which stores metadata and location information, and a set of daemons which manage data movement, deletion, accounting and other functions.

The LDCS system consists of four computing sites with local storage on shared file systems, which are not accessible outside the site, and a GridFTP storage at SLAC, accessible by the collaboration. Simulations run at all four sites, the resulting output files are stored locally and a copy is uploaded through GridFTP to the primary storage LCSC at SLAC. The use case for Rucio in this case is primarily as a catalog of data locations rather than for data transfer. In addition, Rucio provides a metadata catalog of the produced simulation data, using its generic JSON metadata system. This system allows storing and querying arbitrary key-value pairs rather than using a fixed database schema, which for a new experiment allows far greater flexibility in the information it stores. The LDMX simulation software prints names and values of all configurable parameters to a JSON file, which is parsed upon job completion to construct a metadata file with the desired key-value pairs to be stored in Rucio. The database can be queried by end users for job specific information such as walltime, software version used, or output file location, and also for physics related parameters such as simulated beam energy or detector details.

At the end of a simulation job, the output data file is uploaded to the selected LCSC storage by the corresponding A-REX service. When aCT detects that the job and data upload has completed, it downloads the metadata file from A-REX and registers its information along with the location of the new data in Rucio. Files registered in Rucio are guaranteed to exist since registration is only done once the transfer to the final location has succeeded.

\subsection{LDMX Software}

The LDMX software~\cite{Ref:LDMX-SW} is written in C++, based on the Geant4~\cite{Ref:Geant4-1,Ref:Geant4-2}
detector simulation application, and uses ROOT~\cite{Ref:ROOT-1,Ref:ROOT-2} for persistence. The framework uses Python wrappers for both Monte Carlo data generation and subsequent analysis. The software is packaged as a Docker~\cite{Ref:Docker} container with all dependencies, and releases are uploaded to DockerHub. LDCS uses containers converted into Singularity~\cite{Ref:Singularity} images. The images are self contained with no other requirement than having Singularity installed, which reliably defines and freezes the combination of software versions and dependencies used during a certain production run. Shared scripts are distributed to all LDCS sites by use of LDMX's GitHub repository~\cite{Ref:LDMX-SW}. 

In addition to the Singularity image, each site configures two types of runtime environment files (RTEs), also shared through the LDMX GitHub. One sets the path to local storage and specifies if the site should keep a local copy of the output. The other sets a path to a specific Singularity image. 
During job submission, software tags are requested through the desired Singularity image RTE, and only sites with this RTE enabled will be considered by aCT for job submission.
Individual configuration for each site, such as specifics for queuing or usage limits, are set within the A-REX configuration file, and are hidden from the end user.

\subsection{Simulation Workflow}

The outcome in any given event of electrons impinging on the target is probabilistic in nature, and, similarly to many other particle physics experiments, LDMX uses Monte-Carlo based simulations to create an ensemble of predicted event outcomes, given certain beam and detector configurations. The Geant4 framework is used to simulate interactions of particles with the matter in the detector. The simulations of dark matter production use simulated kinematics and probabilities from MadGraph/MadEvent~\cite{ref:madgraph}, embedded in Geant4 for propagation through the detector simulation. 
The software can be used either to generate Monte-Carlo simulated data using input parameters, or to process input files for analysis. Standard simulation jobs process on the order of a million of events each, and typically produce hundreds of megabytes of data in a one-hour job.

\subsection{Monitoring the System}

A web-based dashboard was developed using Prometheus and Grafana, allowing LDMX production managers the overview needed to observe ongoing campaigns and spot potential issues. Metrics on the states of queued and running jobs, as well as Rucio storage information, are periodically collected by aCT and published to a local Prometheus server. A Grafana instance collects and displays this information in a collection of time-series graphs.

\begin{figure}[!h]
\centering
\includegraphics[width=13cm,clip]{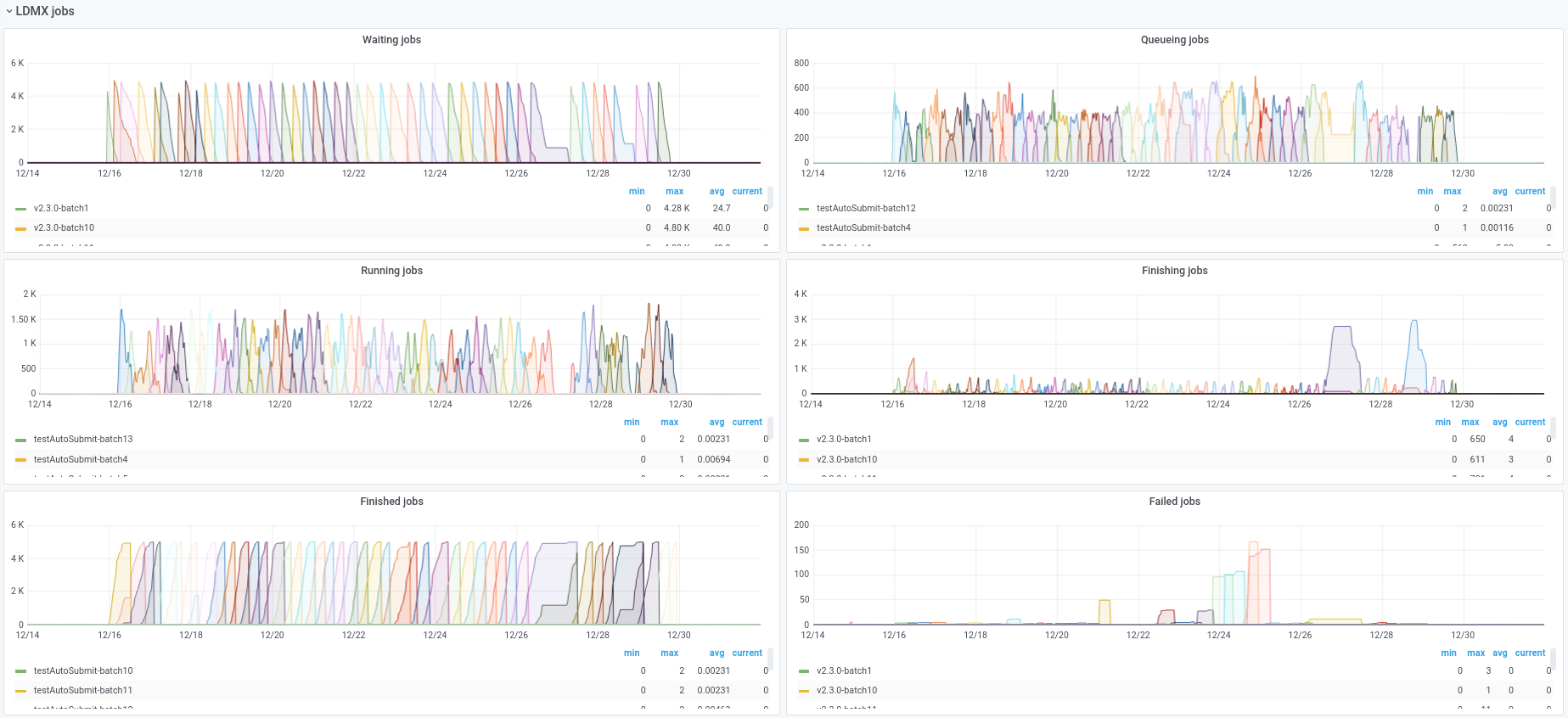}
\caption{Snapshot of the LDCS dashboard, showing number of jobs in different states over time. Different colors are different batches of jobs.}
\label{fig:ldmxdash}
\end{figure}

Figure~\ref{fig:ldmxdash} shows the LDCS dashboard during a simulation campaign in the second half of December 2020. The different colors represent different batches of 5000 jobs which were submitted over time. Each panel shows the number of jobs in each state: waiting to be submitted, queued on a site, running, finishing (waiting for upload of the output data), finished and failed. There were generally between 500 and 1500 jobs running, and this number varied according to the scheduling policies of the batch systems on the sites.

\begin{figure}[h]
\centering
\includegraphics[width=11cm,clip]{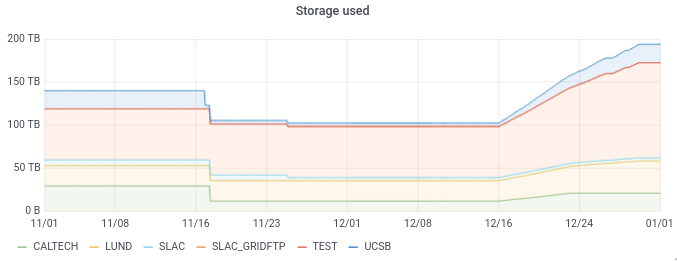}
\caption{Data storage usage over time for LDMX.}
\label{fig:rseusage}
\end{figure}
Figure~\ref{fig:rseusage}, taken from the LDCS dashboard, shows the storage used in each Rucio-managed storage in November and December 2020. In mid-November some data were deleted. A large simulation campaign in the last two weeks of December produced almost 100TB of data. At the end of 2020, Rucio was cataloging a total of almost 200TB of data in 1.7 million files.

\section{Large Scale Test Campaigns}
\label{exp}
The system was exercised over several test campaigns, detailed in Table~\ref{tab:tests}, allowing continuous development and fine-tuning of its capabilities. The main functionalities implemented as a result of tests are described in the following sections.

While the simulation software used during a campaign is identical across sites, each batch system has specific characteristics such as hardware, number of computing nodes, disk space available, and priority rules. The first two campaigns took a few days to run, and highlighted differences in needed walltime between sites, as well as  a need for a higher submission rate of short-duration jobs to the sites more distant to the aCT (physically located in Norway). Two longer campaigns of 225000 1-hour jobs each, lasting a few weeks, more thoroughly explored data transfer and storage needs. 
Site sysadmins reported on average 1h of work, including for site setup, for each test. 
One reason for the low workload for site admins during tests was regular, low-overhead meetings, allowing predictable engagement rather than requiring site admins to be available and responsive at any given moment.

\begin{table}[!h]
\footnotesize
\centering
\begin{tabular}{ p{0.7cm} | p{0.95cm} | p{1.05cm} | p{0.95cm} | p{0.8cm} | p{5.8cm} }
 \textbf{Month}& \textbf{Number of jobs } &  \textbf{Run time} &  \textbf{File size (MB)} &  \textbf{Failure rate} &  \textbf{Most common failure mode } \\ 
\hline
April  & 5k & 6 hours & 500 & 44\% & Walltime limit; Successful job logs filling up monitoring machine disk; Site job cancellation policy \\  
\hline
 May  & 1M & 6 min & 20 & 0.6\%& Job exceeding memory limit    \\
 \hline
 Oct  & 225k & 1 hour & 225 & 1\%& Sync problems upon repeated A-REX restarts \\
 \hline
 Dec  & 225k & 1 hour & 225 & 0.03\%& Singularity script not executing properly  \\    
\end{tabular}
\caption{\raggedright Date, total number of jobs, average run time per job, average output file size, failure rate, and dominant failures modes, for four larger test campaigns in 2020.}
\label{tab:tests}
\end{table}

\subsubsection*{Resubmission}
Exit codes from the batch queue are parsed for known failure modes. Transient job failures (exceeding walltime limits, queue preemption and internet connection glitches) will trigger job resubmission, as the job is likely to succeed at a different site. Jobs exceeding memory limits (which will reoccur for every new attempt) will not be resubmitted. Implementing resubmission rules after the first campaign drastically reduced the failure rate (see Table~\ref{tab:tests}). Resubmission has not yet been implemented for poorly understood cases where Singularity reports that the image is corrupted, despite that it works well on other nodes. 

\subsubsection*{Job bookkeeping and LDCS system metadata}
\begin{figure}[h]
\centering
\includegraphics[width=7.5cm,clip]{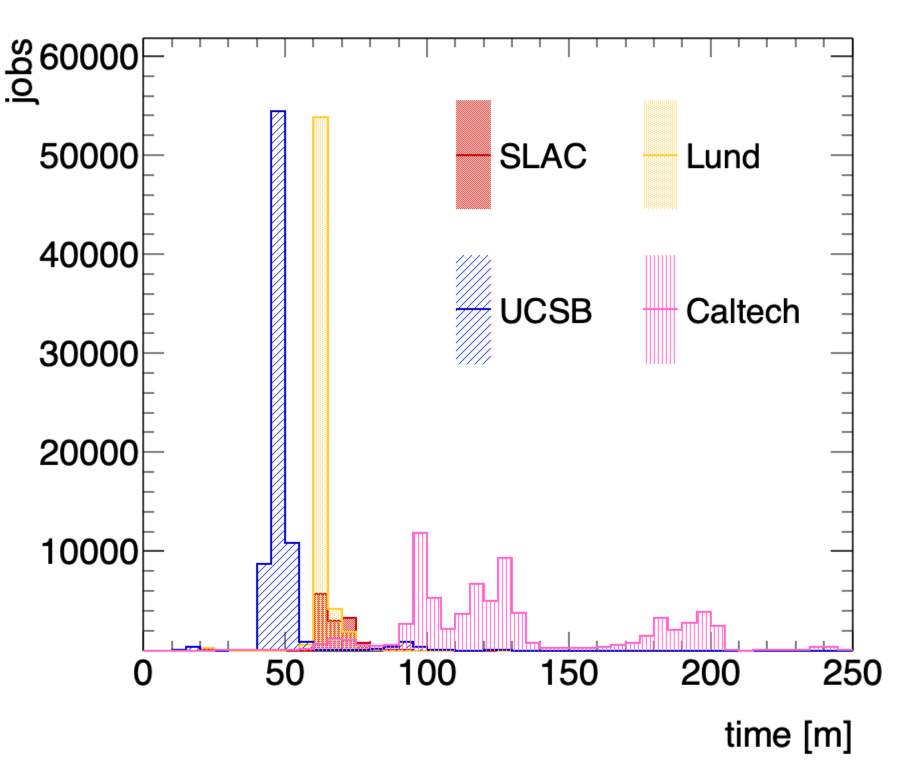}
\caption{Distribution per site of job walltimes for the December 2020 test campaign.}
\label{fig:walltimes}
\end{figure}

Logs of failed jobs are retained, enabling identification of both overarching and site-specific failure modes and exit codes. The Grafana graphical dashboard reveals characteristic time structures in the batch queues. 
Figure~\ref{fig:walltimes} shows the December campaign  job walltimes at each site, extracted from job metadata in Rucio.
Despite identical software and simulation tasks, the spread is significant, due to the range of different CPUs deployed in each center. The job metadata shows rich potential for understanding the individual batch systems. 

\subsubsection*{Data storage and transfer}
The production manager typically redirects all output to a common storage location. 
Site admins control whether a local copy of the job output is kept, with a parameter of the RTE. 
The A-REX transfer system prevents local copies from being deleted until transfer has completed successfully. 
It is also clear that at least 2000 files of O(100) MB can be held and transferred to one single site without the transfer process becoming a significant bottleneck.

\subsubsection*{Submission automation}
To further structure the output, submission of the 225000 jobs  needed for the larger scale campaigns is split into batches. 
Each output file is assigned to a Rucio dataset using the Rucio concept of scope and a unique submission batch name. 
The number of active jobs in the system is monitored in aCT. A periodically run bash submission script on aCT checks if this number has dropped below a set threshold, in which case it submits a new batch, with batch specific parameters updated as needed. This continues until the desired number of batches is reached.
This automation significantly reduces the time spent by the production manager, as well as the risk for errors introduced in manual edits of the job configuration.

\section{Ongoing and Future Work}
\label{future}
Data storage and retention is a perennial problem for large datasets. Work is underway at Lund to create a distributed storage that can be used for intermediate calculations. The final data will be sent to the primary storage at SLAC, to be directly accessible by all members of LDMX, with the option to preserve copies of important data in both storage systems. 
Access to the Lund storage is based on ARC's own GridFTP storage implementation. The performance and data load information collected by this setup will be used to assess which technology best fulfills LDMX distributed data needs in the medium and long term.

Since the larger campaigns in 2020, data-driven job brokering as described in Sec.~\ref{LDCS} has been implemented.
This functionality opens up possibilities of both separate reprocessing of existing simulations, and end user analysis jobs, which will make LDCS a full-fledged distributed computing system. 
Further development based on the LDCS components will enable smaller sites or individual researchers to deploy an LDMX Analysis Station, a computational unit that can perform analysis tasks on pre-fetched datasets (transferred from an LCSC).

Finally, leveraging existing technologies gives the privilege of not having to worry too much about future changes in middleware; most of this work will be done by the component developers and simply adopted by LDCS.

\section{Conclusion}

Particle physics collaborations are global affairs and utilizing the computational resources available, in concert with one another, and with minimal administrative overhead, is a nontrivial endeavor. LDMX is a small scale experiment with nine contributing institutes located  on both sides of the Atlantic, many of them having access to computing facilities. Therefore it was quite a natural choice to start a distributed computing pilot and investigate if such an infrastructure could serve LDMX in the future. The successful deployment of the LDCS  e-infrastructure has shown that a distributed computational infrastructure is both manageable and resource efficient, with  evidence that: 

\begin{itemize}
\item Dividing the IT infrastructure over multiple sites automatically leads to knowledge transfer,
avoiding the situation where IT knowledge is concentrated at a single central site.
\item The LDCS system offers protection against site-lockup, a common mistake when an IT technology is too  customized to a specific site setup, later leading to portability issues.
\item LDCS offers a solution to the possible data storage and data transfer bottlenecks of the centralized approach. Furthermore, LDCS brings protection against data transfer failures due to the automatic transfer retries.
\item Running simulation on LDCS introduced a higher level of automation and reproducibility into the typical LDMX computing workflows, with a standardized software setup using singularity images and the definition of a consistent LDMX metadata structure.
\item With the enhanced reproducibility, LDCS also offers (limited) protection against data losses, with the possibility of regenerating a local dataset on an LCSC site. 
\item LDCS enabled LDMX to take advantage of computing available during natural ebbs and flows of computing workload which occur across the participating computing clusters.
\end{itemize}
The distributed computing pilot program for LDMX  proved the feasibility of building and running a lightweight distributed e-infrastructure for small experiments. Utilizing open-source code with minor modifications and engaging remote sites on a regular basis, the program successfully integrated disparate computational resources into a single compute entity requiring minimal administrative efforts, and seamless to the scientific researcher.

\section*{Acknowledgements}

We would like to thank the Rucio and ARC team for support during the integration of their services into LDCS, and Viktor Shcherbatyuk for LDCS site setup and operation at the California Institute of Technology (Caltech), USA.
The Rucio and aCT services are kindly hosted by the University of Oslo's Center for Information Technology.
In Lund, the computing infrastructure was enabled by resources provided by: the Swedish National Infrastructure for Computing (SNIC)~\cite{SNIC} at LUNARC~\cite{LUNARC}, partially funded by the Swedish Research Council through grant agreement no. 2018-05973; LUNARC's 
own infrastructure and personnel; the IT support and infrastructure at Lund University.
At UCSB, use was made of computational facilities purchased with funds from the National Science Foundation (CNS-1725797) and administered by the Center for Scientific Computing (CSC). The CSC is supported by the California NanoSystems Institute and the Materials Research Science and Engineering Center (MRSEC; NSF DMR 1720256) at UC Santa Barbara.
LKB acknowledges support from the Knut and Alice Wallenberg foundation (no 2018.0429).  RP acknowledges support through the Swedish Research Council (2019-03436) as well as The L’Oréal-UNESCO For Women in Science in Sweden Prize with support of the Young Academy of Sweden.


\begin{thebibliography}{}

\bibitem{wlcg} The LCG TDR Editorial Board, 
LCG-TDR-001, CERN-LHCC-2005-024 (2005) 
 
\bibitem{wlcg2} I. Bird et al., 
LCG-TDR-002, \href{https://cds.cern.ch/record/1695401}{CERN-LHCC-2014-01} (2014)


\bibitem{ldmx} T. Åkesson et al.
 (2018), \href{https://arxiv.org/abs/1808.05219}{arXiv:1808.05219}
  

  
    


\bibitem{LCLS2} LCLS-II Conceptual Design Report (2014), \href{https://portal.slac.stanford.edu/sites/ad_public/people/galayda/Shared_Documents/LCLS-II\%20Conceeptual\%20Design\%20Report.pdf}{LCLSII-1.1-DR-0001-R0} 

\bibitem{atlas} ATLAS Collaboration, JINST \textbf{3} S08003 (2008)

\bibitem{arc} M.~Ellert et al., Future Gener. Comput. Syst. \textbf{23} 219-240 (2007)


\bibitem{act} JK.~Nilsen et al., J. Phys.: Conf. Ser. \textbf{664} 062042 (2015)

\bibitem{rucio} M.~Barisits et al., Comput. Softw. Big Sci. \textbf{3} 11 (2019)

\bibitem{Ref:LDMX-SW} \textit{LDMX software}: \href{https://github.com/LDMX-Software/ldmx-sw/}{https://github.com/LDMX-Software} (2021). Accessed: 2021-02-09


\bibitem{Ref:Geant4-1} Geant4 collab., Instruments and Methods in Physics Research A \textbf{506} 250-303 (2003)

\bibitem{Ref:Geant4-2} Geant4 collab., IEEE Transactions on Nuclear Science \textbf{53} No. 1  270-278 (2006)


\bibitem{Ref:ROOT-1} R.~Brun, F.~Rademakers, Nucl. Instrum. Meth. A \textbf{389} 81-86 (1997)

\bibitem{Ref:ROOT-2} \textit{ROOT}: \href{https://github.com/root-project}{https://github.com/root-project} (2021). Accessed: 2021-02-09

\bibitem{Ref:Docker}
D. Merkel, 
Linux Journal 2014, \textbf{239} 2  (2014)

\bibitem{Ref:Singularity} GM.~Kurtzer et al., PLoS ONE \textbf{12} 5: e0177459 (2017), \href{https://doi.org/10.1371/journal.pone.0177459}{10.1371/journal.pone.0177459}

\bibitem{ref:madgraph}
J. Alwall et al., 
JHEP \textbf{09}, 028 (2007), \href{https://arxiv.org/abs/0706.2334}{arXiv:0706.2334}

\bibitem{SNIC} Swedish National Infrastructure for Computing (SNIC), \href{https://www.snic.se/}{https://www.snic.se/}

\bibitem{LUNARC} LUNARC, Center for scientific and technical computing at Lund University, \href{http://www.lunarc.lu.se/}{http://www.lunarc.lu.se/}

\end{thebibliography}
{\RaggedRight

}

\end{document}